\documentclass[twocolumn]{jpsj2} 
\title{Wang-Landau Simulation for the Quasi-One-Dimensional Ising Model}

\author{Takayuki \textsc{Tanabe}$^{}$ and  Kouichi \textsc{Okunishi}$^{1}$}

\inst{Graduate School of Science, Niigata University, Igarashi 2, 8050 Niigata 950-2181, Japan. \\
$^{1}$Department of Physics, Faculty of Science, Niigata Univeristy, Igarashi 2, 8050 Niigata 950-2181, Japan.}

\abst{We revisit the nature of the quasi-one-dimensional Ising model on the basis of  Wang-Landau simulation.
We introduce the density of states in the two-dimensional energy space corresponding to the intra- and interchain directions.
We then analyze the interchain coupling dependence of specific heat of the anistropic two-dimensional Ising model in the context of the density of states, and further discuss the size dependence of  peak temperature.
We also discuss the feature of the phase transition in the three-dimensional case.
}

\kword{quasi-one-dimension, Wang-Landau simulation, density of states}

\begin{document}
\maketitle

\section{Introduction}
The effect of the interchain coupling in the quasi-one-dimensional(Q1D) spin system has recently attracted much attention. 
In general, the 1D spin system shows no phase transition at a finite-temperature. In the Q1D system, however, the weak interchain coupling induces a finite-temperature phase transition.
In fact, the 3D long-range ordering for the Q1D  system has been a long-standing issue.\cite{q1dreview}
The recent experimental developments enable the precise investigation of 3D ordering in a wide variety of Q1D systems,  such as coupled $S=1/2$ Heisenberg chains.\cite{q1d3}
Moreover, very recently, it has been shown that the Q1D Ising-like XXZ antiferromagnet BaCu$_2$V$_2$O$_8$ exhibits an exotic incommensurate spin order in a magnetic field\cite{bcvo1,bcvo2}, in which the Ising anisotropy plays an essential role.

Motivated by the experimental results above, we reexamine the phase transition of the Q1D Ising system:
\begin{equation}
 \mathcal{H} =-J\sum_{<i,j>_\shortparallel}S_i S_j-J'\sum_{<i,j>_\bot}S_i S_j,
 \label{e1} 
\end{equation}
where $S\in \pm 1$ is the Ising spin variable, $<i,j>_\shortparallel$ indicates spin pairs along the chain direction, and $<i,j>_\bot$ means the pairs perpendicular to the chain.
Thus, $J$ and $ J'$ respectively represent the coupling constants for the intra- and inter-chain  interactions.
Of course the universality of the Ising transition due to the $Z_2$ symmetry breaking itself is independent of the interchain interaction. 
However,  quantitative details of the interchain dependence of the phase transition still involve an interesting problem\cite{fisher1,stanley,graimlandau, kwlee}, which is essential to resolve experimental results.
Recently, the universal reduction in the effective coordination number has also been reported for the Q1D Ising model.\cite{stodo}
The aim of this paper is to understand the role of interchain coupling in the context of the energy density of states (DOS) based on Wang-Landau simulation\cite{wl1,wl2}. The size dependence of the transition temperature of the Q1D Ising system is also discussed. 

For the quantitative analysis of the phase transition of the Q1D system,  recall that the energy scale of interchain coupling is much smaller than that of intrachain coupling, and thus the transition temperature becomes very low;
the conventional metropolis Monte Carlo simulation based on a local spin flip often fails in relaxation to the appropriate equilibrium state.
Recently, an efficient cluster algorithm has been proposed for the Q1D system. \cite{tota}
In this paper, however, we employ such a generalized ensemble method as multicanonical simulation\cite{berg}.
In particular, the Wang-Landau simulation\cite{wl1,wl2} enables us to estimate DOS efficiently through a random walk in the energy space, and to avoid trapping in a metastable state.
Then we can resolve the contribution of typical configurations at low temperatures,  which provides an essential viewpoint of the low-temperature behavior of the Q1D system.

This paper is organized as follows. In the next section, we briefly explain details of the Wang-Landau simulation for the Q1D system.
In particular, we introduce the DOS of two-dimensional energy space for the  intra- and inter-chain directions.
In \S 3, we discuss the phase transition in the 2D case in the context of DOS and then analyze the interchain interaction dependence of the phase transition.
For 3D case, we also discuss the nature of the phase transition.
In \S4,  the summary and discussions are given.

\section{Simulation Details}

The Wang-Landau simulation is based on a random walk in the energy space  without trapping metastable state and enables us to estimate DOS.
For the Q1D system, however,  the energy scale of interchain coupling is fairly different from that of intrachain coupling.
For the purpose of treating such a highly anisotropic system more efficiently, we further introduce the two-dimensional energy space defined by
\begin{equation}
e_{\shortparallel}\equiv \sum_{<i,j>_\shortparallel} S_iS_j ,\quad {\rm and}\quad e_{\bot}\equiv \sum_{<i,j>_\bot} S_iS_j,\label{e2d}
\end{equation}
where $e_{\shortparallel}$ and $e_{\bot}$ respectively denotes the (dimensionless) unbiased energies for the intra- and inter-chain directions.
Then the total energy is given by $E=-Je_{\shortparallel}-J'e_{\bot}$.
The Wang-Landau simulation itself is performed {\it for this two-dimensional space of the spatially isotropic Ising model} and then  DOS $g(e_{\shortparallel},e_{\bot})$ is obtained in $(e_{\shortparallel},e_{\bot})$ space.
The expectation value for various $J'$ can be obtained  through reweighting.
Here, it should be noted that the multicanonical simulation for the two-dimensional parameter space was successfully applied to such complex systems as the polymer system\cite{iba} and the spin glass system\cite{hatano}. 
Moreover, the Wang-Landau sampling drastically enhances the accessibility to the multi-dimensional parameter space in various situations.
An interesting point in the present case may be that the dimensionality of the parameter space is directly related to the spatial dimension.

The detailed conditions for the Wang-Landau simulation are given as the follows.
At the start of simulation,  DOS is unknown, so it is simply set at $g(e_{\shortparallel},e_{\bot})=1$ for all possible  $(e_{\shortparallel},e_{\bot})$ .
Then we begin a random walk in $(e_{\shortparallel},e_{\bot})$ space with a probability proportional to $1/g(e_{\shortparallel},e_{\bot})$ .
The transition probability from $(e_{\shortparallel}^{'},e_{\bot}^{'})$ to $(e_{\shortparallel}^{''} ,e_{\bot}^{''})$ is
\begin{equation}
{\rm Prob}( (e_{\shortparallel}^{'},e_{\bot}^{'}) \to  (e_{\shortparallel}^{''} ,e_{\bot}^{''})
)
= \min \left[   \frac{g(e_{\shortparallel}^{'},e_{\bot}^{'})}
 {g(e_{\shortparallel}^{''} ,e_{\bot}^{''})},1        \right],
\end{equation}
and  $g(e_{\shortparallel},e_{\bot})$ is iteratively
updated by the modification factor $f$  as 
\begin{equation}
 \ln g(e_{\shortparallel},e_{\bot}) \to \ln g(e_{\shortparallel},e_{\bot}) +\ln f,
\end{equation}
when the state is visited. At the same time, the histogram is updated like $H(e_{\shortparallel},e_{\bot}) \to  H(e_{\shortparallel},e_{\bot}) + 1 $. 
When the histogram becomes ``flat", the modification factor is reduced, $ \ln f \to (\ln f) /2 $, and  we reset the histogram  to zero.
Then we  perform a random walk again.
To check the flatness of the histogram, we use the criterion $ (H_{\max}-H_{\min})/(H_{\max}+H_{\min}) \leq 0.1 \sim 0.3 $, where $H_{\max}$ and  $H_{\min}$ are the maximum and minimum histogram counts, respectively.\cite{hklee}
We end the simulation when the modification factor is smaller than a predefined value (we set $\ln f_{final} = 10^{-6} $).
The initial value of the modification factor is $\ln f_0 =1$.
The update of $g(e_{\shortparallel},e_{\bot})$ and $ H(e_{\shortparallel},e_{\bot})$ is performed every $N$ spin flips, where $N$ is the number of spins in the system.\cite{zb}

\section{Results}
\subsection{2D Ising model}
\begin{figure}[t]
\begin{center}
\includegraphics[width=6.0cm]{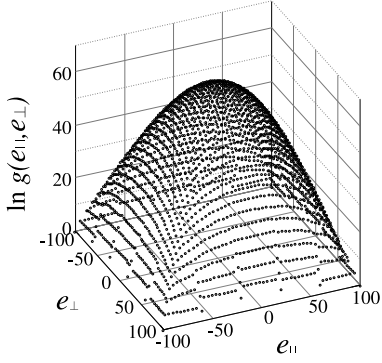}
\end{center}
\caption{Two-dimensional DOS of the 2D Ising model of system size L=10.}
\label{f1}
\end{figure}
\begin{figure}[t]
\begin{center}
\includegraphics[width=6.0cm]{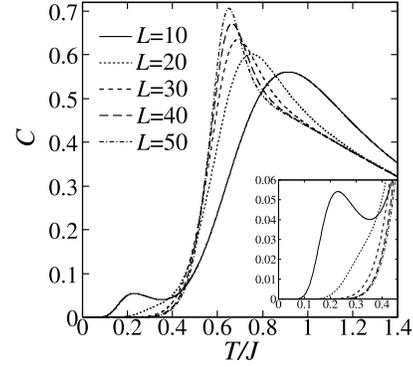}
\end{center}
\caption{Specific heat for $J'/J=0.025$. Inset: magnification of the low-temperature peak.}
\label{f2}
\end{figure}

Let us first consider the 2D Ising model on a $ L \times L $ square lattice, the exact solution of which is well known\cite{onsager} and very useful for verifying the simulation results.
The Hamiltonian of the 2D Ising model is written as
\begin{equation}
 \mathcal{H} =-J\sum^L_{i,j} S_{i,j} S_{i+1,j}
      -J' \sum^L_{i,j} S_{i,j} S_{i,j+1},
\end{equation}
where $i$ and  $j$ are the indexes of  the intra- and inter-chain directions, respectively.
The unbiased energies for the intra- and inter-chain directions are  explicitly given by $e_{\shortparallel}=\sum^L_{i,j}  S_{i,j} S_{i+1,j}$ and $e_{\bot}=\sum^L_{i,j} S_{i,j} S_{i,j+1}$, respectively.
Since $g(e_{\shortparallel},e_{\bot})=g(-e_{\shortparallel},e_{\perp})=g(e_{\shortparallel},-e_{\bot})=g(-e_{\shortparallel},-e_{\bot})$ holds, it is sufficient to perform simulation in the region $e_{\shortparallel}\geqq0,e_{\bot}\geqq0$. 
The system sizes are $L=10,20,30,40$, and $50$.
The maximum histogram count  per stage is $H_{\max}=3024$, and the maximum Monte Carlo steps per stage is $1.7\times 10^8$ for the $L=10$ system.
Then the total number of stages is 21, and the total CPU time is 2 minutes with a 2.66GHz Core2Duo processor.
For $L=30$, $H_{\max}= 7981$, the maximum Monte Carlo steps per stage is $3.3\times 10^{11}$. 
The total number of stages is also 21  and the total CPU time is 55 hours.
In Fig. 1, we show the typical result of DOS $g(e_{\shortparallel},e_{\bot})$ for $L=10$.

On the basis of DOS $g(e_{\shortparallel},e_{\bot})$, we calculate specific heat; 
Figure 2 shows the size dependence of the specific heat for $J'/J=0.025$.
According to the exact solution of the 2D Ising model, the transition temperature for $J'/J=0.025$ is given as $T_c/J=0.6221\cdots$. 
The result clearly shows that the peak of $C$ corresponding to the critical divergence gradually develops for $L=50$ in the vicinity of $T_c$.
In addition to the critical point, we can also see a small peak in the low-temperature region $T/J\sim 0.2$.
As $L$ increases, the peak temperature of this small peak shifts to a higher temperature side and the peak height itself decreases rapidly.

\begin{figure}[t]
\begin{center}
\includegraphics[width=6.0cm]{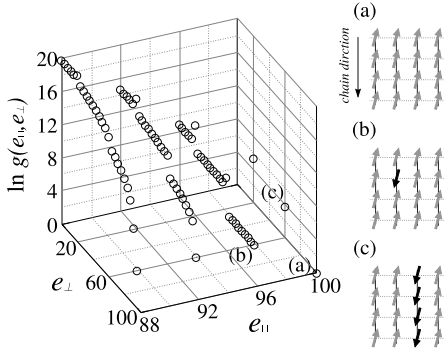}
\end{center}
\caption{Two-dimensional DOS of the 2D Ising model of $ L=10$ near the groundstate($ 88\leqq e_{\shortparallel}\leqq 100, 0\leqq  e_{\bot}\leqq 100$).
The figures in the right panel indicate the configurations for (a) the ground state, (b) single-spin flipped state, and  (c) chain flipped state.}
\label{f3}
\end{figure}
\begin{figure}[t]
\begin{center}
\includegraphics[width=6.0cm]{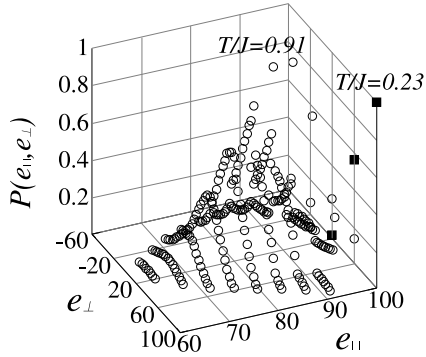}
\end{center}
\caption{Energy distribution function for $J'/J=0.025$ with $L=10$.
The solid squares indicate the distribution function for $T/J=0.23$, which corresponds to the low temperature peak of the specific heat.
The open circles indicate that for the high temperature peak $(T/J=0.91)$.
Each distribution function is normalized so that the maximum value corresponds to unity.}
\label{f4}
\end{figure}

In order to see the origin of the low-temperature peak, we show the DOS $g( e_{\shortparallel},e_{\bot} ) $ in the low-energy region in Fig. 3.
Note that the scale of $e_\shortparallel$ is much smaller than that of $e_\bot$; 
Since  $J'\ll J$ for the Q1D system,  the range of the horizontal axis in Fig. 3 is adjusted by the ratio: $e_\shortparallel/e_\bot\sim J'/J =  1/40$.
The ground state energy is $E_g=-(J+J')L\times L$ and its configuration is illustrated as Fig. 3(a), which is located at $( e_{\shortparallel},e_{\bot} )=(100, 100)$.
As a low-energy excitation,  the single-spin flipped state given by Fig. 3(b) is usually considered, whose energy is given as $E_{(b)}=E_g+4(J+J')$. 
For the Q1D system, however,  another important excitation we should discuss is ``chain flipped excitation'', a typical example of which is depicted in  Fig. 3(c), and its energy is given as $E_{(c)}=E_g+ 4J'L$.
The DOS of the chain flipped configuration is located  at $e_\bot = 60, 20 \cdots $ on the edge of $e_{\shortparallel}=0$.
In Fig. 1,  we can also confirm that the DOS of these configurations at the edges  deviate from the ``bulk'' value in the $( e_{\shortparallel},e_{\bot} ) $ plane.
Moreover, note that the  ``gap'' in DOS  at every $e_\bot=20$ in Fig. 3  also originates from the chain structure of the lattice.
Since $L <  J/J'$, we can see that the chain flipped excitation has a lower energy than the single spin flipped state, but it becomes the predominant excitation at a low temperature.
As $L$ increases beyond $J/J'$, the energy of the chain flipped excitations shifts to the higher-energy region, so that the contribution of such configurations decreases gradually.
Thus the low-temperature peak of specific heat in Fig. 2 can be well described by the chain flipped excitations, which is peculiar to the Q1D system.

In order to see the weight of each energy state in the equilibrium,  we calculate the energy distribution function 
\begin{equation}
P \left( e_{\shortparallel},e_{\bot}   \right)= g( e_{\shortparallel},e_{\bot} )\exp\left[( Je_{\shortparallel}+J' e_{\bot} )/T\right]
\end{equation}
for $J'/J=0.025$, which is shown in Fig. 4.
Note that $T/J=0.23$ is the temperature of the low-temperature peak of specific heat and $T/J=0.91$ corresponds to the high-temperature peak of $L=10$.
In the figure, the predominant contribution at $T/J=0.23$ clearly comes from the states at the edge of $e_{\bot} =100$, which implies that the chain flipped state is essential for the small peak;
For a relatively small system size ($L<J/J'$), the energy of the excitations actually satisfies $4(J+J') > 4J'L$.
On the other hand, $P( e_{\shortparallel},e_{\bot} )$ for  $T/J=0.91$ shows a Gaussian-like shape at approximately  $( e_{\shortparallel},e_{\bot} )\sim (90,20)$, where the chain flipped state gives only a minor contribution in DOS.

As mentioned above, the predominant contribution to the low-temperature peak($T/J=0.23$) is the chain flipped states near the ground state.
This implies that the polarization of the spins in the same chain is basically frozen and the aligned spins in the chain can behave as a single spin, which forms an effective 1D spin chain through the weak interchain coupling $LJ'$ in the interchain direction.
Thus, we can see that the fluctuation in the interchain direction is predominant for the low-temperature peak, while for the high temperature peak, the fluctuations in both the intra- and inter-chain directions  give the significant contributions.
Of course, the low temperature peak is basically a finite-size effect and vanishes in the bulk limit.
However,  the region where the finite-size effect can clearly appear is up to $L\sim J/J'$, which is a certain large number for the Q1D system. 
This implies that the true critical divergence of specific heat is eventually masked by the analytic contribution of the low-temperature peak, up to $L\sim J/J'$.
Thus, the finite-size scaling analysis based on the data  $L<J/J'$ should be performed carefully.\cite{fss}

In order to extract the proper critical behavior for $L<J/J'$, we examine the decomposition of specific heat into three parts: the fluctuation along the chain, $C_\shortparallel$; the fluctuation in the interchain  direction, $C_\bot$; and cross term of the intra- and inter-chain directions, $C_{\rm cross}$.
\begin{eqnarray}
 C&=&(\langle E^2 \rangle - \langle E \rangle^2 )/ NT^2 \nonumber \\
  &=& C_{\shortparallel}+C_{\bot}+C_{\rm cross},
\\
C_{\shortparallel}&=& J^2 ( \langle e_{\shortparallel}^2 \rangle  - \langle e_{\shortparallel}  \rangle ^2 )/ NT^2, \\
C_{\bot}&=& J'^2 ( \langle e_{\bot}^2 \rangle  - \langle e_{\bot}  \rangle ^2 )/ NT^2 ,\\
C_{\rm cross}&=& 2J J' \left( \left< e_{\shortparallel} e_{\bot}   \right>
-\left< e_{\shortparallel} \right> \left< e_{\bot} \right> \right)
/N T^2  .
\end{eqnarray} 
The intra- and inter-chain fluctuations basically reflect the 1D nature of the spin fluctuations. 
Thus, for $C$, $C_\shortparallel$, and $C_\bot$, we have to see the critical fluctuation in the background of the predominant 1D behavior.
On the other hand, the cross term $C_{\rm cross}$ exhibits no such 1D behavior and thus we can expect more direct observation of the critical fluctuation.

\begin{figure}[t]
\begin{center}
\includegraphics[width=8cm]{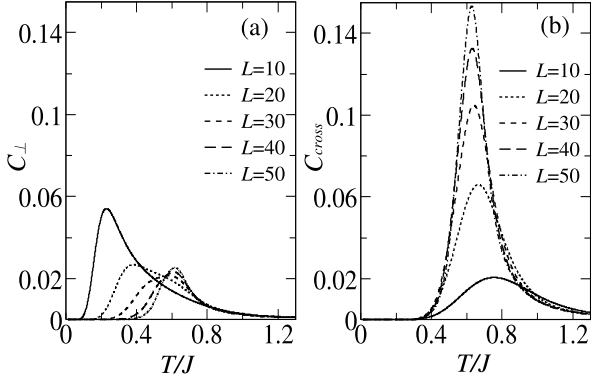}
\end{center}
\caption{Decomposed specific heats for $J'/J=0.025$: (a) the energy fluctuation in the interchain direction, $C_{\bot}$, and the cross term of the chain and  interchain directions, $C_{\rm cross}$.}
\label{f5}
\end{figure}
\begin{figure}[t]
\begin{center}
\includegraphics[width=8.0cm]{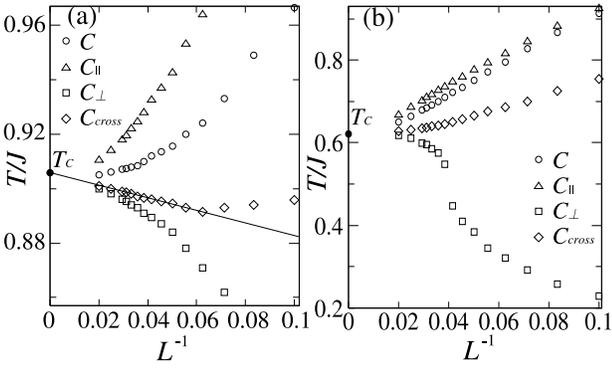}
\end{center}
\caption{Size dependences of the peak temperatures for $C$, $C_{\shortparallel}$,  $C_{\bot}$, and $C_{\rm cross}$: (a) $J'/J=0.1$ and (b) $J'/J=0.025$.
The solid circles at the vertical axis indicate the exact transition temperatures.}
\label{f6}
\end{figure}

In  Fig. 5, we show $C_{\bot}$ and $C_{\rm cross}$ for $J'/J=0.025$($C_{\shortparallel}$ is not presented here). 
In the figure, $C_{\bot}$ shows a Schottky-like peak for small system sizes ($L< 20$).
As $L$ increases, the peak position shifts to the high-temperature side and the peak height itself rapidly decreases.
This behavior is consistent with the fact that the interchain fluctuation is the predominant contribution for $L<J/J'$.
Indeed, we have verified that the low temperature peak of $L=10$ can be well fitted by the specific heat of the 1D Ising chain of the effective coupling $LJ'$ with $L=10$.
In addition to the low-temperature peak, a broad peak also emerges at approximately $T/J\sim 0.6$, as $L$ increases; this peak corresponds to the critical divergence in the bulk limit.
Thus, the  crossover of $C_\bot$ from the effective 1D Ising model behavior to the 2D Ising model clearly appears at $L\sim J/J'$.
On the other hand, the cross term $C_{\rm cross}$ shows the divergence behavior only near the correct critical temperature $T_c=0.622\cdots$.
This suggests that $C_{\rm cross}$  captures the critical behavior more effectively than the total specific heat $C$.

We further analyze the size dependence of the  peak temperatures of the decomposed specific heats $C$, $C_{\shortparallel}$,  $C_\bot$, and $C_{\rm cross}$.
Let us write the peak temperature for the system size $L$ as $T_c(L)$.
Then, in the critical regime, peak temperature is expected to follow the size scaling
\begin{equation}
T_c(L)-T_c=A L^{-1/\nu} \label{shiftfss},
\end{equation} 
where $A$ is a nonuniversal constant.
First we discuss the results for $J'/J=0.1$,  which are illustrated in Fig. 6(a).
In the figure, we can see that $C$, $C_{\shortparallel}$, and  $C_\bot$ gradually approach $T_c$, for which no scaling behavior is  observed.
However, the cross term of the specific heat  $C_{\rm cross}$ well satisfies eq. (\ref{shiftfss}) within a relatively small system size ($1/L<0.06$), suggesting that $C_{\rm cross}$ is more effective for capturing the critical behavior than $C$.
Another interesting feature of the Q1D system is that $T_c(L)$ approaches $T_c$ from the under side of $T_c$, namely,  $A>0$,  for sufficiently large $L$.
This behavior is contrasted to $A < 0 $ in the isotropic case where the peak temperatures of all the $C$s monotonically approach $T_c$ from $T>T_c$.

The peak temperatures for $J'/J=0.025$, which are shown in Fig. 6(b), demonstrate a more typical size dependence of the Q1D system;
The peak of $C_\shortparallel$ monotonically decreases from the upper side of $T_c$, while $C_\bot$ clearly exhibits the crossover behavior.
For small $L$($<0.04$), the peak position of $C_\bot$ originates from the chain flip configuration.
However, we can see that it rapidly crossovers to that of the critical behavior at $1/L\sim 0.04 $.
On the other hand, $C_{\rm cross}$ seems to approach $T_c$ smoothly,  suggesting that $C_{\rm cross}$ is more suitable for finite size analysis of the critical behavior.
For $J'/J=0.025$, however, note that system size may be still insufficient for the precise verification of the critical exponent $\nu$.

\subsection{3D Ising model}

Let us discuss the 3D Ising model along the same line of argument as that of the 2D case. 
The Hamiltonian is written as
\begin{eqnarray}
 \mathcal{H}&=&-J \sum^L_{i,j,k} S_{i,j,k} S_{i,j,k+1}\nonumber \\
 &-& J'  \sum^L_{i,j,k} [ S_{i,j,k} S_{i+1,j,k}+ S_{i,j,k} S_{i,j+1,k}  ],
\end{eqnarray}
where $k$ is  assumed to run in the chain direction.
In actual computations, the maximum histogram count per stage is $H_{\max}=6024$, and the maximum Monte Carlo steps per stage is $6.5\times 10^9$ for $L=6$ system.
The total number of stages is 21 and the total CPU time is 50 minutes.
For $L=10$, $H_{\max}= 6748$, the maximum Monte Carlo steps per stage is $7.5\times 10^{11}$, and the total CPU time is about 6 days.
Here, we note that, for the 3D Ising model, Wang-Landau simulation in the 2D energy space (2) sometimes does not achieve good convergence near the edges of the 2D energy space.
In such a case, we have additionally performed Wang-Landau simulation for the conventional 1D energy space to obtain specific heats.

Figure 7(a) shows the size dependence of the specific heat $C$ for $J'/J=0.025$ up to $L=18$. 
In the figure,  we can see the broad maximum of $C$ of $L=6$ at approximately $T/J\sim 1.0$. 
At the same time,  a small peak emerges in the low-temperature region, reflecting the 1D nature of the system.
 As $L$ increases,  this small peak rapidly merges with a broad peak coming down from the higher-temperature side. 
We can then see that the merged peak rapidly develops into a sharp peak associated with the critical divergence at $T/J\sim 0.8$.

\begin{figure}[h]
\begin{center}
\includegraphics[width=8cm]{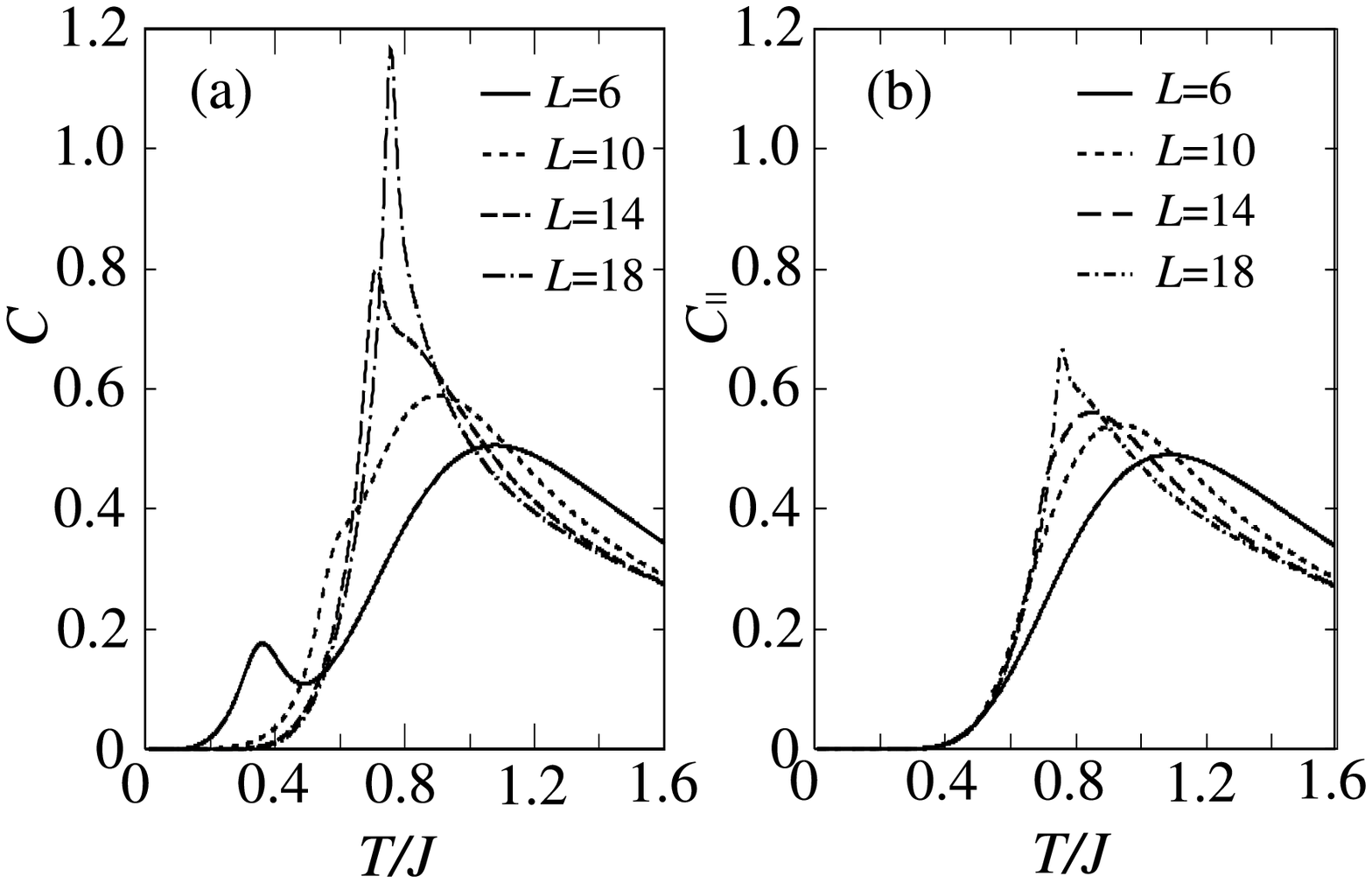}
\includegraphics[width=8cm]{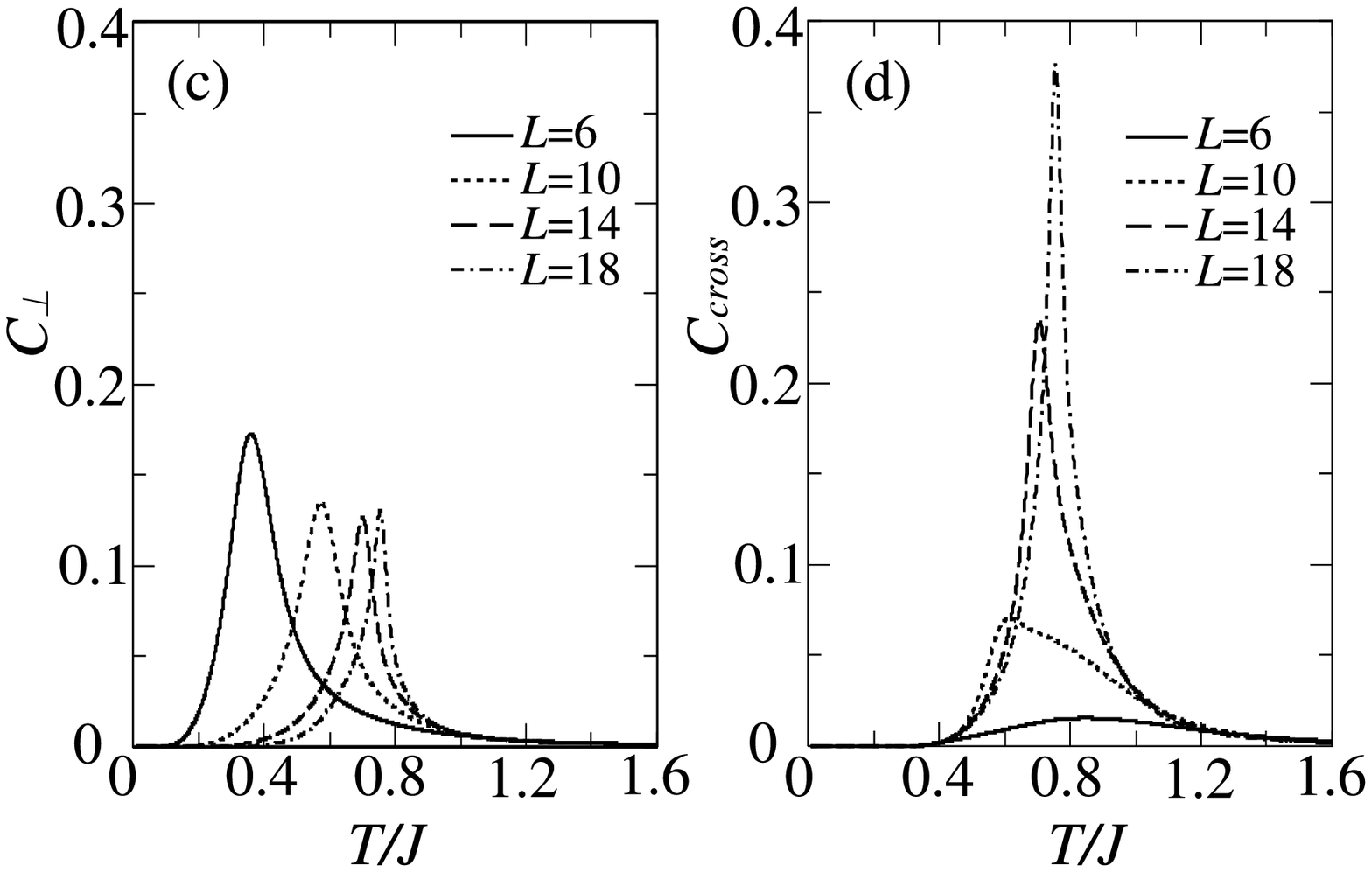}
\end{center}
\caption{ $C$, $C_{\shortparallel}$, $C_{\bot}$, and $C_{\rm cross}$ for the 3D Ising model of $J'/J=0.025$.}
\label{f7}
\end{figure}

We next resolve the peak structure of the total specific heat using $C_{\shortparallel}$, $C_{\bot}$, and $C_{\rm cross}$.
Figure 7(b) illustrates $C_{\shortparallel}$,  which exhibits a broad peak for a small system size.  
Since $J\gg J'$, this broad peak can be attributed to the spin fluctuation in the chain, which is governed by the energy scale $J$.
As $L$ increases, the peak temperature decreases to $T\sim 0.8$ and the peak height itself increases in accordance with the criticality.
On the other hand, $C_{\bot}$ in Fig. 7(c) shows a clear finite-size peak originating from the interchain fluctuation of the energy scale $LJ'$. 
As $L$ increases, the peak temperature gradually increases from $T/J\sim 0.4$ to $0.8$.
Then, an interesting point about $C_{\bot}$ is that  the shape of the peak is almost unchanged during shifting,  in contrast to the 2D case where the peak considerably reduces its shape.
Here, let us recall that, at sufficiently low temperatures, the effective  spins are frozen in the chain direction form the 2D network.
Thus, an important difference between the 2D and 3D cases is that, for 3D, the effective 2D Ising model in the small $J'$ limit can involve the quasi-critical divergence, while for 2D, the specific heat of the effective 1D Ising model shows no  such divergence since there is no phase transition in the 1D Ising model.
In Fig. 7(d), we finally show $C_{\rm cross}$, the peak of which  develops near $T_c$ and is smoothly connected to the critical divergence.

In Fig. 8, we summarize the above size dependences of the peak temperatures for the specific heats.
In the figure, the horizontal axis indicates the scaled system size $L^{-1/\nu}$, where we have used $\nu=0.6301 $.\cite{hwjblote}
We can see that the round peak of $C_{\shortparallel}$ comes from  the higher temperature side, but it still does not reach the scaling region.
On the other hand, we can see that  $C_{\bot}$ and $C_{\rm cross}$ are well fitted by linear functions, which are respectively shown as solid and broken lines in Fig.8.
The straightforward extrapolation yields $T_c\simeq 0.85$, which is consistent with the precise estimation $T_c=0.834$ based on the simulation up to the size $10\times 10\times 80$ (the result is not presented here).
A similar analysis of susceptibility was also reported in Ref. \citen{kwlee}, where the peak temperature of the intrachain spin fluctuation behaves similarly to $C_\bot$.
The present result is consistent with the previous analysis of susceptibility\cite{kwlee}.
As can be seen Fig. 7(a), the divergences of $C_{\rm cross}$ and $C_\bot$ massively contribute to the critical divergence of the total specific heat $C$ within a small system size.
This suggests that $C_{\rm cross}$ can be expected to be suitable for the finite-size analysis of the critical behavior as well, although $C$ exhibits a rather complicated size dependence of the peak structure in the 3D case.

\begin{figure}[h]
\begin{center}
\includegraphics[width=7.0cm]{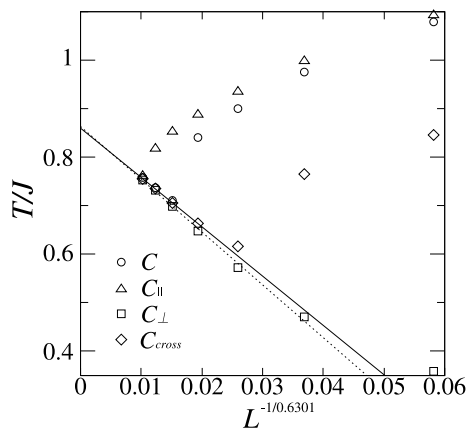}
\end{center}
\caption{Size dependences of the peak temperatures of $C$, $C_{\shortparallel}$, $C_{\bot}$, and $C_{\rm cross}$ for the 3D Ising model of $J'/J=0.025$.
The scale of the horizontal axis follows $L^{-1/\nu}$ with $\nu =0.6301 $. \cite{hwjblote}}
\label{f8}
\end{figure}

\section{Summary and Discussion}

We have studied the feature of the Q1D Ising model using Wang-Landau simulation.
In order to treat the difference between  the energy scales for the intra- and inter-chain directions, we have particularly introduced the two-dimensional energy space ($e_{\shortparallel}, e_{\bot}$) corresponding to the intra- and inter-chain directions.
We further decomposed the total specific heat $C$ into  contributions from intrachain fluctuation,  $C_{\shortparallel}$, interchain fluctuation, $C_{\bot}$, and the cross term of the intra- and inter-chain fluctuations, $C_{\rm cross}$.
Then the finite-size effect peculiar to the Q1D system is discussed on the basis of the two-dimensional DOS, and it was demonstrated that the chain flip configuration plays an essential role for the low-temperature peak of specific heat.
We have also analyzed the shift exponent of the peak of the specific heat, and then found that $C_{\rm cross}$ can capture the critical behavior  more effectively than the total specific heat $C$,  within a relatively small system size.
We have also discussed the qualitative difference between 2D and 3D cases in the low-temperature and small-$J'$ limits;
in the 2D case, the crossover of $C_\bot$ from an effective 1D chain to a 2D model occurs rapidly at $L\sim J/J'$.
In the 3D case, the effective 2D Ising model itself involves the quasi-critical divergence, because $C_\bot$ smoothly crossovers from the effective 2D model into the 3D critical behavior.

In this paper, we have analyzed the Q1D Ising model in the context of the two dimensional DOS.
The actual computational cost for obtaining the two-dimensional DOS increases rapidly, with increasing system size, and then the cluster algorithm seems to be more efficient for a simulation of a larger system.
However, the present description based on the two-dimensional DOS provides the essential insight for the qualitative understanding of the low-energy excitations in the Q1D system.
In addition, to suppress the finite-size effect peculiar to the Q1D system, the aspect ratio usually follows the ratio of anisotropic correlation length in analyzing the critical behavior.\cite{aspect}
We can also see that a possible aspect ratio of the system is $ L'/L =(D-1)J'/J$, where $D$ is the dimension of the system, $L$ is the length of a chain, and $L'$ is the size of the interchain direction.
This is because the scale of the single-spin flip excitation and the chain flipped state can be on the same order $4J+4(D-1)J' \simeq 4(D-1)J'L$  for $J\gg J'$.

\section*{Acknowledgments}

This work is supported by  Grants-in-Aid for Scientific Research from the Ministry of Education, Culture, Sports, Science and Technology of Japan (Nos. 18740230 and 20340096), and  by a Grant-in-Aid for Scientific Research on Priority Area ``High-Field Spin Science in 100T''.
One of the authors (K.O.) would like to thank T. Suzuki and M. Kikuchi for valuable discussions.


\begin{thebibliography}{99} 
\bibitem{q1dreview} L. J. de Jongh, and A. R. Miedem: Adv. Phys. {\bf 23} (1974) 1. 


\bibitem{q1d3}I. Tsukada, Y. Sasago, K. Uchinokura, A. Zheludev, S. Maslov, G. Shirane, K. Kakurai, and E. Ressouche: Phys. Rev. B {\bf 60} (1999) 6601.
 
\bibitem{bcvo1}
S. Kimura, T. Takeuchi, K. Okunishi, M. Hagiwara, Z. He, K. Kindo, T. Taniyama, M. Itoh: Phys. Rev. Lett. {\bf 100} (2008) 057202;
S. Kimura, M. Matsuda, T. Masuda, S. Hondo, K. Kaneko, N. Metoki, M. Hagiwara, T. Takeuchi, K. Okunishi, Z. He, K. Kindo, T. Taniyama, and M. Itoh: Phys. Rev. Lett. {\bf 101} (2008) 207201.

\bibitem{bcvo2} K. Okunishi and T. Suzuki: Phys. Rev. B {\bf 76} (2007) 224411; 
T. Suzuki, N. Kawashima, and K. Okunishi: J. Phys. Soc. Jpn. {\bf 76}, (2007) 123707.

\bibitem{fisher1}C. Y. Weng, R. B. Griffiths and M. E. Fisher: Phys. Rev. {\bf 162} (1967)475.;  M. E. Fisher: Phys. Rev. {\bf 162} (1967) 480.


\bibitem{stanley} L. L. Liu and H. E. Stanley: Phys. Rev. Lett. {\bf 29} (1972)  927;  L. L. Liu and H. E. Stanley: Phys. Rev. B {\bf 8} (1973)  2279.

\bibitem{graimlandau} T. Graim and D. P. Landau: Phys. Rev. B {\bf 24} (1981) 5156.

\bibitem{kwlee} K. W. Lee: J. Phys. Soc. Jpn. \textbf{71}  (2002) 2591


\bibitem{stodo} S. Todo: Phys. Rev. B. \textbf{74} (2006) 104415.

\bibitem{tota} T. Nakamura: Phys. Rev. Lett. {\bf 101} (2008) 210602.

\bibitem{berg} B. A. Berg and T. Neuhaus: Phys. Rev. Lett. {\bf 68} (1992) 9.


\bibitem{wl1} F. Wang and D. P. Landau: Phys. Rev. Lett.  \textbf{86} (2001) 2050.
\bibitem{wl2} F. Wang and D. P. Landau: Phys. Rev. E. \textbf{64} (2001) 056101.

\bibitem{iba}Y. Iba, G. Chikenji, and M. Kikuchi: J. Phys. Soc. Jpn. {\bf 67} (1998) 3327.

\bibitem{hatano} N. Hatano and J. E. Gubernatis: Prog. Theor. Phys. Suppl. {\bf 138} (2000) 442.

\bibitem{hklee} H. K. Lee, Y. Okabe,  and D. P. Landau: Bull. Comput. Phys. Commun. \textbf{175}  (2006) 36.
\bibitem{zb} C. Zhou and R. N. Bhatt: Phys. Rev. E. \textbf{72} (2005)
025701(R).

\bibitem{onsager} L. H. Onsager: Phys. Rev. {\bf 65} (1944) 117.

\bibitem{fss} In order to extract the proper critical behavior, the aspect ratio of the system may be adjusted to the Q1D system. 
In this paper, however, we discuss the size dependence within  square or cubic systems.

\bibitem{hwjblote} H. W. J. Bl$\ddot{\rm o}$te, E. Luijten, and J. R. Heringa: J. Phys. A \textbf{28}  (1995) 6289


\bibitem{aspect} K. Binder and J.-S. Wang, J. Stat. Phys {\bf 55 } (1989) 87. 
See also ref. \citen{graimlandau}





\end{thebibliography}
\end{document}